\documentstyle[prb,aps,floats,psfig]{revtex}
\begin{document}
\draft
\twocolumn
\wideabs{
\title{Scaling analysis of the magnetic-field-tuned quantum
transition in superconducting amorphous In--O films}
\author{V.F.~Gantmakher\thanks{e-mail: gantm@issp.ac.ru},
M.V.~Golubkov, V.T.~Dolgopolov, G.E.~Tsydynzhapov, and A.A.~Shashkin}
\address{Institute of Solid State Physics, Russian Academy of
Sciences, 142432 Chernogolovka, Russia}
\maketitle

\begin{abstract}
We have studied the magnetic-field-tuned superconductor-insulator
quantum transition (SIT) in amorphous In--O films with different
oxygen content and, hence, different electron density. While for
states of the film near the zero-field SIT the two-dimensional
scaling behaviour is confirmed, for deeper states in the
superconducting phase the SIT scenario changes: in addition to the
scaling function that describes the conductivity of
fluctuation-induced Cooper pairs, there emerges a
temperature-dependent contribution to the film resistance. This
contribution can originate from the conductivity of normal
electrons.
\end{abstract}
}

The scaling analysis is an important experimental tool for studying
quantum phase transitions. For two-dimensional (2D) disordered
superconductors, alongside with the zero-field
superconductor-insulator transition (SIT) as driven by disorder
change in the film, there exists a SIT that is induced by a normal
magnetic field. A scenario of the field-induced SIT was proposed in
Ref.~\onlinecite{Fisher}: at zero temperature the normal magnetic
field alters the state of a disordered film from superconducting at
low fields, through a metallic one at the critical field $B=B_c$ with
the universal sheet resistance $R_c$ close to $h/4e^2\simeq
6.4$~k$\Omega$, to an insulating state at fields $B>B_c$.  The SIT
was supposed to be continuous with the correlation length $\xi$ of
quantum fluctuations, diverging as $\xi\propto(B-B_c)^{-\nu}$, where
the critical index $\nu>1$. At a finite temperature the size of
quantum fluctuations is restricted by the dephasing length
$L_\phi\propto T^{-1/z}$ with the dynamical critical index $z$, which
determines the characteristic energy $U\sim\xi^{-z}$ and

is  expected to be equal to $z=1$ for SIT.

The ratio of these two length parameters defines the
scaling variable $u$ so that near the transition point ($T=0,B_c$)
all data $R(T,B)$ as a function of $u$ should fall on a universal
curve
\begin{equation}
R(T,B)\equiv R_cr(u),\qquad u=(B-B_c)/T^{1/z\nu}.\label{x=}
\end{equation}
Although small in the scaling region, temperature dependent
corrections with the leading quadratic term are expected to the
critical resistance $R_c$ \cite{Fisher,QPT}.

The above theoretical description is based on the concept of electron
pair
localization which has been supported by a recent publication
\cite{Larkin}. In that paper it is shown that for 2D superconducting
films with sufficiently strong disorder the region of fluctuation
superconductivity, where the localized
electron pairs

(called also boson \cite{Fisher} and cooperon \cite{Larkin})

occur, should
extend down to zero temperature. In this region the unpaired
electrons are supposed to be localized because of disorder in a film.

So far, a theory of the field-driven 3D quantum SIT has not been
created. An idea to consider the quantum SIT for 3D disordered
systems in zero magnetic field in terms of charged boson localization
\cite{golda} was not at first accepted because the fluctuation
superconductivity region was regarded to be small. In fact, as was
shown later in Ref.~\onlinecite{Bul}, the fluctuation region enlarges
as the edge of single electron localization is approached. This gives
an opportunity to apply the scaling relation deduced for 3D boson
localization \cite{Fish2} also for the field-induced SIT description

\begin{equation}
R(T,u)\sim T^{-1/z}\tilde r(u), \label{e3D}\end{equation}
where $\tilde r(u)$ is a universal function and the scaling variable
$u$ is assumed to have the same form as defined by Eq.~(\ref{x=}).

From Eqs.~(\ref{x=}) and (\ref{e3D}) it follows that in the vicinity
of $B_c$ the isotherms $R(B)$ are straight lines with slopes

\begin{equation}
\frac{\partial R}{\partial B}\propto T^{-(d-2+1/\nu)/z},
\label{sca} \end{equation}

where $d$ is the system dimensionality.
Because the behaviours of the resistance in the relations (\ref{x=})
and (\ref{e3D}) are very different, the problem of the film
dimensionality is of major importance.

Data obtained in experimental studies on $a$-In--O \cite{HP},
$a$-Mo--Ge \cite{Kapit}, and $a$-Mo--Si \cite{Okuma} followed the 2D
scaling relation (\ref{x=}) except for the universality of the $R_c$
value.

This was regarded as evidence of existence of SIT. The
failure in satisfying the scaling relations in ultrathin Bi films
\cite{cher} was interpreted as indication of the absence of SIT
and crossover observation between different flux-flow regimes.
Studies \cite{HP,Kapit,Okuma} did not give arguments backing boson
localization.
At the first time such arguments are appeared by
interpretation of  the resistance drop at high fields observed on
$a$-In--O films
\cite{gg2,JETPL}.

Here, we perform the detailed study of the scaling relations near the
field-induced SIT for different states of an $a$-In--O film. We find
that the 2D scaling relation (\ref{x=}) holds for film states near
the zero-field SIT but progressively fails as the zero-field SIT is
departed from. This failure is manifested by the appearance of an
extra temperature-dependent term in the film resistance.

The experiments were performed on 200 \AA\ thick amorphous In--O
films evaporated by e-gun from high-purity $In_2 O_3$ target onto a
glass substrate \cite{InO}.

This
material proved to be very useful for
investigations of the transport properties near the SIT
\cite{HP,HP1,gg2,ShOv,Kim}. Oxygen deficiency compared to fully
stoichiometric insulating compound In$_2$O$_3$ causes the film
conductivity. By changing the oxygen content one can cover the range
from a superconducting material to an insulator with activated
conductance \cite{ShOv}. The procedures to change reversibly the film
state are described in detail in Ref.~\onlinecite{gg2}. To reinforce
the superconducting properties of our films we used heating in vacuum
up to a temperature from the interval 70 -- 110$^\circ$C until the
sample resistance got saturated. To shift the film state in the
opposite direction we made exposure to air at room temperature. As
the film remains amorphous during these manipulations, it is natural
to assume that the treatment used results mainly in a change of the
total carrier concentration $n$ and that there is a critical
concentration $n_c$ corresponding to the zero-field SIT.

The low-temperature measurements were carried out by a four-terminal
lock-in technique at a frequency of 10~Hz using two experimental
setups: a He$^3$-cryostat down to 0.35~K or Oxford TLM-400 dilution
refrigerator in the temperature interval 1.2~K -- 30~mK. The ac
current was equal to 1~nA and corresponded to the linear regime of
response. The aspect ratio of the samples was close to one.

We investigated three different homogeneous states of the same
amorphous In--O film \cite{rem}. We characterize the sample state by
its room temperature resistance $R_r$. Assuming that the disorder for
all states is approximately the same, we have for the carrier density
$n\propto 1/R_r$, i.e., the smaller $R_r$, the deeper the state in the
superconducting phase and, hence, the larger the value of $B_c$.

\begin{table}

\caption{Parameters of the studied states of the sample.}
\begin{tabular}{c|cccc}
State&$R_r$, k$\Omega$&$R_c$, k$\Omega$&$B_c$, T&$\alpha$, K$^{-1}$\\
\tableline
1&3.4&7.8&2.2&0\\
2&3.1&8&5.3&-0.1\\
3&3.0&9.2&7.2&-0.6\\
\end{tabular}
\label{t1}
\end{table}

The parameters of the investigated states are listed in
Table~\ref{t1}. State 1 is the closest to the zero-field SIT and
state 3 is the deepest in the superconducting phase.

\begin{figure}
\psfig{file=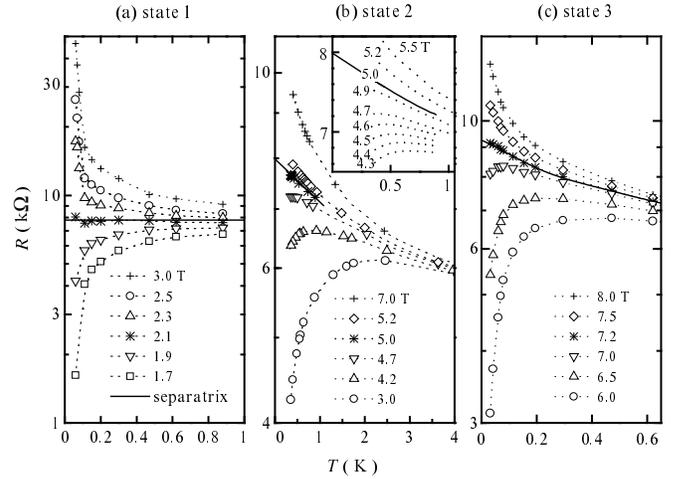,width=\columnwidth,clip=}
\smallskip\caption{Temperature dependences of the resistance of the studied
states for different magnetic fields. The separatrices $R_c(T)$ are
shown by solid lines.}

\label{f1}

\end{figure}

Sets of the isomagnetic curves $R(T)$ for all studied states are
depicted in Fig.~\ref{f1}. For each set the curves can be divided
roughly into two groups by sign of the second derivative: the
positive (negative) sign corresponds to the insulating
(superconducting) behaviour. Henceforth, the boundary isomagnetic
curve $R_c(T)$ between superconductor and insulator, which
corresponds to the boundary metallic state at $T=0$, will be referred
to as separatrix. While for state 1 it is easy to identify the
horizontal separatrix in accordance with Eq.~(\ref{x=}), for states 2
and 3 the fan and separatrix  are
"tilted", i.e., each of the curves in the
lower part of the fan is a maximum at a temperature $T_{\rm max}$
which shifts with $B$. To determine the separatrix $R_c(T)$ one has
to extrapolate the maximum position to $T=0$ for which it is good to
know the extrapolation law as the accessible temperature range is
restricted.

\begin{figure}
\psfig{file=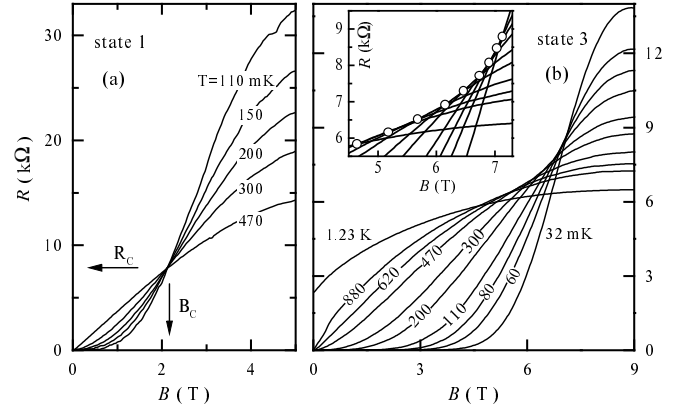,width=\columnwidth,clip=}

\smallskip\caption{Isotherms in the ($B,R$) plane for states 1 and 3. The curve
intersection region for state 3 is blown up in the inset. The circles
mark the crossing points of the isotherms with neighbouring
temperatures.}

\label{f2}

\end{figure}

The absence of a horizontal separatrix for states 2 and 3 can also be
established from the behaviour of isotherms $R(B)$ (Fig.~\ref{f2}).
As seen from the figure, the isotherms of state 1 cross at the same
point ($B_c,R_c$) whereas those of state 3 form an envelope.

\begin{figure}
\centerline{\psfig{file=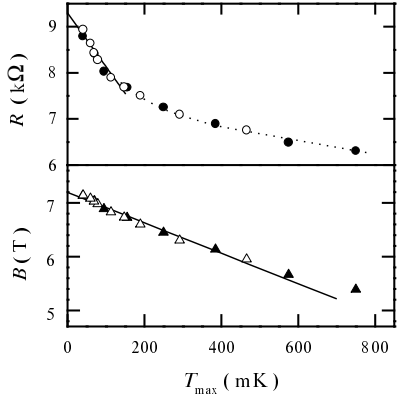,width=0.6\columnwidth,clip=}}

\smallskip\caption{Dependences $R(T_{\rm max})$ and $B(T_{\rm max})$ as
determined from the data in Fig.~\protect\ref{f1}b (open symbols) and
Fig.~\protect\ref{f2}b (filled symbols). The values of $R_c$ and
$B_c$ are obtained with the help of linear extrapolations (solid
lines). The dotted line is a guide to the eye.}

\label{f4}

\end{figure}

To determine $B_c$ and $R_c$ for states 2 and 3 we use the simplest
linear extrapolation to $T=0$ of the functions $R(T_{\rm max})$ and
$B(T_{\rm max})$, see Fig.~\ref{f4}. The open symbols correspond to
the maximum positions on isomagnetic curves (Fig.~\ref{f1}) and the
filled symbols represent the data obtained from the intersections of
consecutive isotherms \cite{mit} (Fig.~\ref{f2}):
if two consecutive isotherms
at close temperatures $T_1$ and $T_2$ intersect at a point
($B_i,R_i$), the isomagnetic curve at the field $B_i$ reaches its
maximum $\approx R_i$ at $T_{\rm max}\approx (T_1+T_2)/2$.

 As seen
from Fig.~\ref{f4}, the dependence $B(T_{\rm max})$ is weak and so we
believe that the linear extrapolation is good to extract $B_c$. In
contrast, the accuracy of the determination of $R_c$ is poor.

\begin{figure}
\centerline{\psfig{file=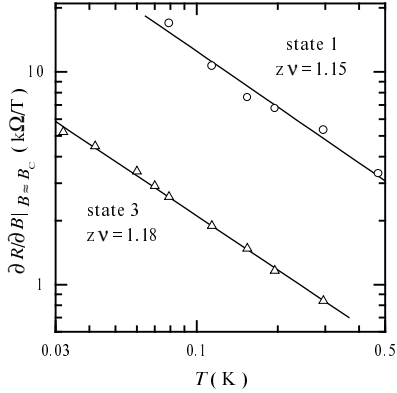,width=0.6\columnwidth,clip=}}

\smallskip\caption{Behaviour of $\partial R/\partial B$ with temperature for
states 1 and 3. The values of exponent $z\nu$ are indicated.}

\label{f3}

\end{figure}

The derivative $\partial R/\partial B$ near $B_c$ as a function of
temperature is shown in Fig.~\ref{f3}.
The exponent turns out to be the same within experimental uncertainty
for the film states 1 and 3 and is in agreement with results of
Refs.~\onlinecite{HP,Kapit} where authors argued observation of the
field-induced 2D SIT for states close to the zero-field SIT. This
fact is in favour of 2D SIT scenario also for deeper film states in
the superconducting phase.

\begin{figure}
\psfig{file=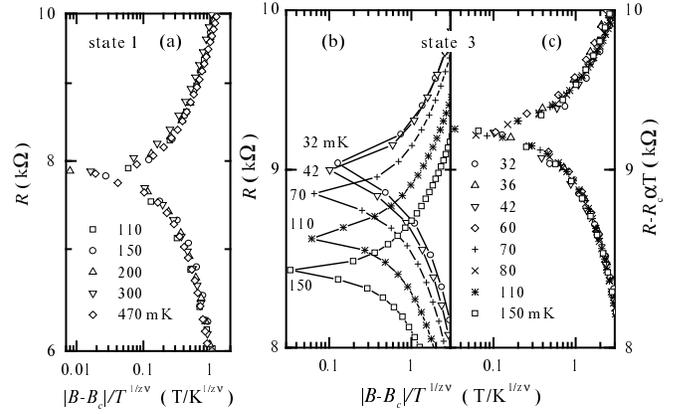,width=\columnwidth,clip=}

\smallskip\caption{Scaling plots for state 1 (a) and for state 3 without (b)
and with (c) the linear temperature term.}

\label{f5}

\end{figure}

Knowing $B_c$ and the scaling exponent we can replot the experimental
data as a function of scaling variable $u$ (Fig.~\ref{f5}). As seen
from Figs.~\ref{f5}a and b, for state 1 the data collapse onto a
single curve whereas for state 3 we obtain a set of similar curves
shifted along the vertical axis. Subtracting formally from $R(T,B)$
the linear temperature term $R_c\alpha T$ (where $\alpha$ is a
factor) does reveal the 2D scaling behaviour for state 3
(Fig.~\ref{f5}c). We note that the procedure of dividing the
experimental data in Fig.~\ref{f5}b by $R_c(T)$, which corresponds to
the formula (\ref{e3D}) for 3D scaling, does not lead to success.

Thus, we find that the 2D scaling holds for states near the
zero-field SIT while the data for deeper states in the
superconducting phase are best described by the relation (\ref{x=})
with an additive temperature-dependent correction $f(T)$

\begin{equation}
R(T,B)\equiv R_c[r(u)+f(T)]. \label{f}
\end{equation}

To get a basis for the formal analysis of the experimental data we
have to answer two questions: (i) whether our film is really 2D; and
(ii) what is the physical origin for the temperature dependence of
$R_c(T)$? In the first case we need to compare the film thickness $h$
with characteristic lengths. These are the coherence length
$\xi_{sc}=c\hbar/2eB_{c2}l$ (where $l$ is the mean free path in
normal state) in superconducting state and the dephasing length
$L_\phi(T)\simeq\hbar^2/m\xi_{sc}T$ \cite{Fisher,QPT} that restricts
the diverging correlation length $\xi$ in the vicinity of quantum
SIT. Knowing the normal state film resistance $R\approx 5$~k$\Omega$
at $T\approx 4$~K and assuming that we deal with the amorphous 3D
metal in which the mean free path is normally close to the lowest
possible value $l\approx 1/k_F$, we estimate the length
$l\approx 8$~\AA. If we crudely evaluate the field $B_{c2}$ at $B_c=7.2$~T
as
determined for state 3, we get

upper limit of

$\xi_{sc}\sim 500$~\AA\ and
$L_\phi\sim 400$~\AA\
at $T=0.5$~K. This supports the 2D scenario of
quantum SIT although in the normal state the film turns out to be 3D.

With respect to the temperature-dependent $R_c(T)$, at finite
temperatures the conductivity of the film near $B_c$ should include
the contribution from localized normal electrons in addition to the
conductivity defined by the diffusion of fluctuation-induced Cooper
pairs \cite{Larkin,Kapit}. It is the normal electron conductivity
that explains the non-universality of the critical resistance
\cite{Kapit} as well as the additional term in Eq.~(\ref{f}). We
write this term in the general form because the linear extrapolation
used is likely to break in the vicinity of $T=0$.

So, all of the experimental observations can be reconciled with the
2D scaling scenario. Intriguingly, the same scaling behaviour has
been established in a parallel magnetic field \cite{gg5}. Although
not in favour of 2D concept, this fact indicates that the
restrictions imposed by the theory \cite{Fisher} may be too severe.

We would like to mention an alternative way to make up for the term
$f(T)$ in Eq.~(\ref{f}): to introduce the temperature-dependent field
$B_c(T)$ defined through the constancy of $R_c$. Formally both ways
are equivalent and correspond to shifts of the isotherms in
Fig.~\ref{f2} either along the $R$-axis or along the $B$-axis so that
in the vicinity of transition a common crossing point is attained. In
contrast to the normal behaviour of the critical fields in
superconductors, the so-defined $B_c(T)$ increases with temperature.
This can be interpreted in terms of temperature-induced boson
delocalization.

In summary, in experiments on amorphous In--O films with different
oxygen content we have found a change of the field-driven 2D SIT
scenario as the film state departs from the zero-field SIT. For deep
film states in the superconducting phase, in addition to the
universal function of scaling variable that describes the
conductivity of fluctuation-induced Cooper pairs, there emerges a
temperature-dependent contribution to the film resistance. This
contribution can be attributed to the conductivity of normal
electrons.

We gratefully acknowledge useful discussions with V.~Dobrosavljevich
and A.I.~Larkin. This work was supported by Grants RFBR 99-02-16117
and RFBR-PICS 98-02-22037 and by the Programme "Statistical Physics"
from the Russian Ministry of Sciences.


\begin{references}
\bibitem{rem} Observation of so-called quasireentrant states for the
field-driven SIT was reported in Refs.~\onlinecite{HP,Kim,Gold} and
explained by inhomogeneities and single-particle tunneling between
superconducting grains \cite{Gold}. This interpretation is supported
in our experiments by the fact that the quasireentrant behaviour
observed in some film states disappeared as a result of annealing the
sample in vacuum for several additional hours after its resistance
had already saturated. We do not discuss quasireentrant states in the
present paper.
\bibitem{mit} Similar extrapolation procedure for determining $R_c$
has been used in Ref.~\onlinecite{mit1} where metal-insulator
transition in 2D system was studied and the carrier density was used
as the driving parameter.
\end{references}
\end{document}